\documentclass[showpacs,preprintnumbers,amsmath,amssymb,prl]{revtex4}
\input epsf.sty

\def\aut#1{#1}
\def\ep{\epsilon}
\usepackage{graphicx}% Include figure files
\usepackage{dcolumn}% Align table columns on decimal point
\usepackage{bm}% bold math
\def\comment#1{}

\newcommand{\vp}{\phi}
\newcommand{\al}{\alpha}
\newcommand{\ra}{\rightarrow}
\newcommand{\be}{\begin{eqnarray}}
\newcommand{\ee}{\end{eqnarray}}
\newcommand{\sgm}{\sigma}
\newcommand{\bt}{\beta}
\newcommand{\dlt}{\delta}
\newcommand{\om}{\omega}
\newcommand{\gm}{\gamma}
\newcommand{\qq}{q}

\begin{document}

\title{Highly Accurate Critical Exponents
from Self-Similar Variational Perturbation Theory}

\author{H. Kleinert$^*$ and V.I. Yukalov$^{*,*}$}

\email{
kleinert@physik.fu-berlin.de;
yukalov@thsun1.jinr.ru}

\affiliation{$^*$Institut f\"ur Theoretische Physik,
Freie Universit\"at Berlin, Arnimallee 14, D-14195 Berlin, Germany}
\affiliation{$^{**}
$Bogolubov Laboratory of Theoretical Physics,
Joint Institute for Nuclear Research, Dubna 141980, Russia}

\pacs{74.20.De,74.25.Fy,74.40.+k}

\vspace{1cm}

\begin{abstract}
We extend field theoretic variational perturbation theory by
self-similar approximation theory, which greatly accelerates
convergence. This is illustrated by re-calculating the critical
exponents of $O(N)$-symmetric $\vp^4$ theory. From only three-loop
perturbation expansions in $4- \epsilon $ dimensions we obtain
{\em analytic results\/} for the exponents, with practically the
same accuracy as those derived recently from ordinary
field-theoretic variational perturbational theory to seventh order.
In particular, the theory explains the best-measured exponent
$\al\approx-0.0127$ of the specific heat peak in superfluid helium,
found in a satellite experiment with a temperature resolution of
nanoKelvin. In addition, our analytic expressions reproduce also
the exactly known large-$N$ behaviour of the exponents $ \nu $
and $ \gamma= \nu (2- \eta ) $ with high precision.
\end{abstract}

\maketitle

\section{I. Introduction}

The precise calculation of critical exponents of phase transitions is an
important theoretical task. On the one hand, these exponents provide us
with basic information on the behaviour of thermodynamic quantities in
the vicinity of critical points. On the other hand, such calculations
require the development of new mathematical techniques to master the
resummation problem of divergent perturbation expansions (see e.g.
\cite{b1,b2}). The comparison of the calculated exponents with experiment
serves as a test for the validity and accuracy of the mathematical methods.

Recently, a powerful method has been developed by one of the authors
called {\em field-theoretic variational perturbation theory} \cite{b3,b4,b5},
which converts divergent weak-coupling into convergent strong-coupling
expansions. This method presents a more powerful alternative to the
previously used method of Pad\'e-Borel resummation. The higher power stems
from an explicit knowledge of the convergence behavior expressed in
terms of the Wegner exponents $\omega$ of approach to the strong-coupling
limit. The higher accuracy has been amply demonstrated by calculating the
critical exponents \cite{b2,b4,b6}, in particular, by predicting the
most-accurately known exponent $\al\approx-0.0127$ \cite{b5} of the specific
heat peak in superfluid helium, found in a satellite experiment with a
temperature resolution of nanoKelvin.

An important feature of any resummation method is the convergence of the
renormalized sequence of approximants. This can be studied by considering
the analytic properties of the sought function with respect to coupling
and by involving the corresponding dispersion relations \cite{b2,b7}. The
field-theoretic variational perturbation theory was shown to possess
an exponentially fast convergence (see the detailed proof in \cite{b8}).

In an independent development,  the other of the authors has developed a
completely different  general resummation scheme called {\em self-similar
approximation theory} \cite{b9,b10,b11,b12,b13}. Also this method
exhibits a fast convergence, which has been demonstrated for a variety
of problems in quantum mechanics, statistical physics, and mathematical
finance (see review-type papers \cite{b14,b15}).

The aim of the present joint paper is to combine the two approaches.
The combination is expected to have the fastest convergence so far.
Indeed, we shall find, that the combined method is capable of reproducing
from three-loop expansions the above mentioned seven-loop result
of field-theoretic variational perturbation theory for critical exponents,
including the most sensitive critical exponent $\alpha$.

In Section II, we give a brief reminder of the basic formulas of
field-theoretic variational perturbation theory, which will be used as
a basis for a further acceleration of the convergence via self-similar
approximation theory to be reviewed in Section III. In Section IV we
develop the combination of the two methods, which is then applied in
Section V to calculate the critical exponents of the $O(N)$-symmetric
$\vp^4$ field theory.

\section{II. From Weak to Strong Coupling}

Physical quantities of interest are usually derived from theories as
divergent series in powers of some bare coupling constant $g_B$. These
provide us with reliable results only for very small $g_B$. Critical
phenomena, however, take place at infinitely large $g_B$ in comparison
with the mass, the inverse length scale of the fluctuations \cite{b2}.
In order to overcome this difficulty one has to reorganize the divergent
weak-coupling series into a convergent strong-coupling expansion.
Such a reorganization is provided by the field-theoretic variational
perturbation theory \cite{b3,b4,b5,b8},  briefly summarized in this section
to recall the principal formulas needed in what follows.

Consider a real function $f(g_B)$ of a real $g_B$, whose limit
$f(\infty)$ we want to find from a divergent weak-coupling expansion
up to order $L$,
\be
\label{1}
f^{(L)}(g_B) = \sum_{n=0}^L f_n\; g_B^n \qquad (g_B\ra 0) \; ,
\ee
with $L=1,2,3,\ldots$ enumerating the maximally available order.
Our aim is to find the behaviour of $f(g_B)$ at $g_B\ra\infty$.
The expansion coefficients, grow factorially with $n$ so that the
series (\ref{1}) could have sense only for very small $g_B$.

Field-theoretic variational perturbation theory is based on the
introduction  in (\ref{1}) of a variational parameter $K$ by the
identical replacement
\be
\label{2}
g_B \rightarrow  \frac{g_B}{(K^2+g_B r)^{q}}\; , \qquad~~~r\equiv
\frac{1-K^2}{g_B},
\ee
where $q$ is a parameter related to the critical Wegner exponents
$\omega $, which in the renormalization group approach to critical
phenomena governs approach to scaling. The parameter $q$ in this
paper corresponds to $q/2$ in the original work \cite{b3}. After the
replacement, the series (\ref{1}) is re-expanded in powers of $g_B$
at fixed $r$, and at the end $r$ is again replaced by $(1-K^2)/g_B$.
This procedure introduces an artificial dependence on the dummy
parameter $K$ which is fixed by searching for a plateau in $K$ which
becomes flatter and flatter for increasing order. The plateau is
horizontal only for the correct choice of $q$, and this condition
will determine the Wegner coefficient $\omega$ \cite{b16}.

In the upcoming calculations, we shall work with a slightly different
but completely equivalent replacement
\be
\label{3}
g_B\rightarrow  \frac{s}{(1-g_Br)^\qq } \; , ~~~~~r\equiv
\frac{\sigma}{g_B}\; ,
\ee
where $\sigma$ is defined as the function of $g_B$:
\be
\label{4}
\sgm \equiv  \sigma (g_B)\equiv 1 -\left (
\frac{s}{g_B}\right )^{1/\qq } \; .
\ee
Following the rules of field theoretic variational perturbation
theory, we have to form the functions
\be
\label{5}
F^{(L)}(g_B,s,\qq ) \equiv f^{(L)}\left
( \frac{s}{(1-g_Br)^\qq }\right )
\ee
to be calculated with the prescription that the  terms  $g_B^n\equiv
{s^n}/{(1-g_Br)^{n\qq }}$ in the truncated series (\ref{1}) are re-expanded
systematically in powers of $g_B$ up to $g_B^{L-n}$. After this, we replace
again $r\rightarrow\sigma /g_B$ and optimize the resulting function in
the variational parameter $s$. Using the binomial expansion
\be
\label{6}
(1-\sgm)^p \simeq \sum_{m=0}^{L-n} C_m^p\; (-\sgm)^m \; , \qquad
C_m^p \equiv \frac{\Gamma(p+1)}{\Gamma(m+1)\Gamma(p-m+1)} \; ,
\ee
we obtain explicitly
\be
\label{7}
F^{(L)}(g_B,s,\qq ) = \sum_{n=0}^L\;
\sum_{m=0}^{L-n} C_m^{-n\qq } \; (-\sgm)^m \; f_n\; s^n \; .
\ee
This must be optimized in $s$, yielding order-dependent function
$s^{(L)}(g_B)$ and an associated $\sigma^{(L)}(g_B)\equiv 1 -
\left({s^{(L)}(g_B)}/{g_B}\right )^{1/\qq }$.

Note that Eqs. (2) and (3) are identities and not directly related to
the functions appearing in the Symanzik-type transformations \cite{b17},
in spite of a certain similarity.

Our aim is to find the behaviour of Eq. (\ref{7}) in the strong-coupling
limit $g_B\ra\infty$. From Eqs. (\ref{3}) we observe that $\sgm\ra 1$
as $g_B\ra\infty$ since, as we shall see, the optimal $s$ is finite.
This allows us to calculate from  Eq. (\ref{7}) as finite approximant
$F^{(L)}(\infty,s,\qq )$. In what follows, we limit ourselves to the
approximants of third order, since then all calculations can be done
analytically. The first three approximants are explicitly
\begin{eqnarray}
F^{(1)}(\infty,s,\qq )& = &f_0 + f_1\; s \; , \label{8}\\
F^{(2)}(\infty,s,\qq )& = &F^{(1)}(\infty,s,\qq ) +\qq \;
f_1\; s +f_2\; s^2 \; , \label{9}\\
F^{(3)}(\infty,s,\qq ) &=& F^{(2)}(\infty,s,\qq ) + \frac{1}{2}\;
\qq  (1+\qq )\; f_1 \; s + 2\qq \; f_2\; s^2 + f_3\; s^3 \; .
\label{10}
\end{eqnarray}

The optimal values  $s^{(L)}=s^{(L)}(\infty)$ are  found
either by extremization
\be
\label{11}
\left.\frac{\partial}{\partial s }\; F^{(L)}(\infty,s,\qq )
\right|_{s=s^{(L)}} = 0 \; ,
\ee
or, when the latter has no real solutions, from the turning points
\be
\label{12}
\left.\frac{\partial^2}{\partial s^2 }\; F^{(L)}(\infty,s,\qq )
\right|_{s=s^{(L)}} = 0 .
\ee
To second order, there exists an extremum at
\be
\label{13}
s^{(2)} = - (1+\qq )\; \frac{f_1}{2f_2} \; .
\ee
To third order, there can be two possibilities. There is an optimal
extremum $s=s^{(3)}$ at one of the roots of the cubic equation
\be
\label{14}
3f_3\; s^2 + 2(1+2\qq )\; f_2\; s + \frac{1}{2}\;
(1+\qq )(2+\qq )\; f_1 = 0 \; ,
\ee
and an turning point at
\be
\label{15}
s^{(3)} = - (1+2\qq )\; \frac{f_3}{3f_3} \; .
\ee
Usually, conditions (\ref{11}) and (\ref{12}) yield optimal values of
$s^{(L)}$ alternatively for odd and even orders $L$, respectively
(see \cite{b2,b3,b4,b5,b6,b8,b18}), and this will be the case in
the upcoming applications of this paper.

After determining $s^{(L)}$, we obtain the optimized approximants
\be
\label{16}
F^{(L){\rm \,opt}}(\infty,\qq ) \equiv F^{(L)}(\infty,s^{(L)},\qq ) \; .
\ee
To second order this is
\be
\label{17}
F^{(2){\rm \,opt}}(\infty,\qq ) = f_0 - (1+\qq )^2\; \frac{f_1^2}{4f_2} \; ,
\ee
and to third order, with $s^{(3)}$ of Eq. (\ref{15}) (see Ref. \cite{b4}),
\be
\label{18}
F^{(3){\rm \,opt}}(\infty,\qq ) = f_0 - (1+\qq )(1+2\qq )(2+\qq )\;
\frac{f_1f_2}{6f_3} + (1+2\qq )^3 \; \frac{2f_2^3}{27f_3^2} \; .
\ee

For each approximant (\ref{16}), we must also specify the parameter $\qq $.
If the Wegner exponent were known from other sources, we could use this.
Otherwise we must determine it order by order, which yields an $L$-dependent
result $\qq =\qq ^{(L)}$, so that the final approximants will be
\be
\label{19}
f^{(L)\rm \,opt} \equiv F^{(L)\rm \,opt}(\infty,\qq ^{(L)}) \; .
\ee

The determination of $\qq ^{(L)}$ proceeds as follows. If we expect the
function $f(g_B)$ to be finite in the strong-coupling limit, which is the
case for the critical exponents, then the logarithmic derivative, to be
referred to as $\beta$-function \cite{b2},
\be
\label{20}
\bt(g_B) \equiv \frac{d\log f(g_B)}{d\log g_B}
\ee
must tend to zero for $g_B\rightarrow \infty$: $\bt(\infty)=0$. From the
expansion (\ref{1}) it is straightforward to derive
\be
\label{21}
\bt^{(L)}(g_B) = \sum_{n=0}^L \beta_n\; g_B^n \; .
\ee
Depending on whether $f_0$ is nonzero or zero,
the coefficients $\beta_n$, up to third order, are given either by
\be
\label{22}
\beta_0 = 0 \; , \qquad \beta_1 = \frac{f_1}{f_0} \; , \qquad
\beta_2 =2\; \frac{f_2}{f_0}\; - \; \frac{f_1^2}{f_0^2} \; ,
\ee
\be
\label{23}
\beta_3 = \frac{f_1^3}{f_0^3} \; - 3\; \frac{f_1f_2}{f_0^2} + 3\;
\frac{f_3}{f_0} \qquad (f_0\neq 0) \; ,
\ee
or by equations
\be
\label{24}
\beta_0 = 1 \; , \qquad \beta_1 = \frac{f_2}{f_1} \; , \qquad
\beta_2 =2\; \frac{f_3}{f_1}\; - \; \frac{f_2^2}{f_1^2} \; ,
\ee
\be
\label{25}
\beta_3 = \frac{f_2^3}{f_1^3} \; - 3\; \frac{f_2f_3}{f_1^2} + 3\;
\frac{f_4}{f_1} \qquad (f_0 = 0) \; .
\ee
Now we treat the expansions $\beta^{(L)}(g_B)$ in the same way as before
$f^{(L)}(g_B)$. We form the approximants $B^{(L)\rm \,opt}(\infty,\qq )$
similar to the way of deriving Eq. (\ref{16}). This is set
$B^{(L)\rm \,opt}(\infty,\qq )$ equal to zero to ensure $\beta(\infty)=0$
in each approximation. This determines the proper parameters $\qq=\qq^{(L)}$.
For instance,
\be
\label{26}
\qq ^{(2)} = 2\; \sqrt{\frac{\beta_0\beta_2}{\beta_1^2}}\; - 1 \; .
\ee

Note that the logarithmic derivative for determining $\qq^{(L)}$ can
be formed from any function of $g_B$ with a constant strong-coupling limit
\cite{b19} , i.e., from any critical exponent, not just from the function
$f(g_B)$ we want to resum at the moment. Usually, the  function $g_R(g_B)$
relating $g_B$ to the renormalized coupling constant $g_R$ is most convenient,
since it is known to highest order.

\section{III. Self-Similar Approximation Theory}

Self-similar approximation theory \cite{b9,b10,b11,b12,b13} is based on
constructing a sequence of optimized approximants, which contain instead
of a variational parameter a {\em trial function}. The general idea of
deriving convergent sequences of optimized approximants with the help of
trial control functions has been suggested in \cite{b20}. The first step
in the optimization procedure is reminiscent of the Euler-Lagrange
variational method. But while the latter is a single-step procedure
\cite{b21}, the optimized perturbation theory runs via a sequence of
better and better approximants.

In the last section we have shown how to calculate a sequence of trial
functions $\{ F^{(L)}(g_B,s,\qq)\}$ by field-theoretic variational
perturbation theory [see Eq. (\ref{7})]. From these $s$ and $\qq $ can
be determined as  functions of $g_B$ by optimization. In self-similar
approximation theory \cite{b9,b10,b11,b12,b13}, the approximants of
different order are considered as a flow on the manifold of approximants,
in which order $L$ of the approximation plays the role of a discretized
pseudotime. In this interpretation, the sequence of approximations
behaves like a dynamical system. The higher approximations will be obtained
by improving the entire optimized functions $s^{(L)}(g_B)$, even if we
are only interested in the strong-coupling value $f(\infty)$, for which
the previous method required only an optimal parameter $s^{(L)}(\infty)$,
Thus we have to perform  the optimization procedure for all $g_B$ {\it before}
going to the limit $g_B\rightarrow \infty$. Instead of Eqs. (\ref{11}) and
(\ref{12}), we have to solve the full extremality condition
\be
\label{27}
\left.\frac{\partial}{\partial s }\; F^{(L)}(g_B,s,\qq )
\right|_{s=s^{(L)}(g_B)} = 0 .
\ee
and, if this has no real solution, the turning point condition
\be
\label{28}
\left.\frac{\partial^2}{\partial s^2 }\; F^{(L)}(g_B,s,\qq
)\right|_{s=s^{(L)}(g_B)} = 0 ,
\ee
to find the lowest approximation for the  trial functions
$s^{(L)}=s^{(L)}(g_B)$. More explicitly, we could  also record the
parameter $q$ at which the optimization is done in the arguments and
write the solution as $s^{(L)}(g_B,\qq)$. But we shall refrain from doing
so to avoid cluttering the notation. For the same reason we shall omit,
for a while, the argument $\qq$ in $F^{(L)}(\infty,s,\qq )$.

Starting from the trial functions $s^{(L)}(g_B)$, we construct an approximation
following the general scheme developed in \cite{b9,b10,b11,b12,b13,b14,b15}.
We define the reonomic function $g_B=g_B^{(L)}(\vp)$ by the reonomic constraint
\be
\label{29} \!\!\!\!\!\!\!\!\!
F^{(1)}(g_B,s^{(L)}(g_B)) =\vp \; ,
\ee
where $F^{(1)}$ is the lowest nontrivial function in the sequence
$\{ F^{(L)}\}$. Now further define an entire sequence of functions
\be
\label{30}
y^{(L)}(\vp) \equiv F^{(L)}(g_B^{(L)}(\vp),s^{(L)}(g_B^{(L)}(\vp)) \; ,
\ee
with the initial term $y^{(1)}(\vp)=\vp$. The set of all functions
$y^{(L)}(\vp)$ for  $L=1,2,3,\ldots$ constitutes a space ${\cal Y}\subset
y^{(L)}(\vp,\qq )$ called {\em approximation space}. The pseudotime
evolution in this space forms a group of self-similarity transformations:
\be
\label{31}
y^{(L+p)}(\vp ) = y^{(L)}(y^{(p)}(\vp) ) \; .
\ee
The property of self-similarity (\ref{31}) guarantees the existence
of a fixed point $y^{*}=y^{(L)*}$ \cite{b14,b15}, which has the property
\be
\label{32}
y^*=y^{(L)}(y^*) \; .
\ee
More explicitly, the fixed point satisfies
\be
\label{33}
y^{(L)*} \equiv F^{(L)*}(g_B^{(L)} ) \equiv
F^{(L)*}(g_B^{(L)},s^{(L)}(g_B^{(L)})) \; ,~~~~L\geq1.
\ee
It defines the desired self-similar approximant
\be
\label{34}
f^{(L)*}(g_B) \equiv F^{(L)*}(g_B(\vp),s^{(L)}(g_B(\vp))) \; .
\ee

In order to find the fixed point we define a  pseudovelocity of the
approximation sequence by the finite difference
\be
\label{35}
v^{(L)}(\vp ) \equiv F^{L+1}(g_B^{(L)}( \vp   ),s^{(L)}( \vp   ) )
- F^{(L)}(g_B^{(L)}(\vp),s^{(L)}(\vp) ) \; .
\ee
If the
$\{ y^{(L)}\}$  with discrete $L=0,1,2,\ldots$ were a flow function
$\{ y^{(t)}\}$ of a continuous time $t\geq 0$, it would follow
a time evolution equation
\be
\label{36}
\frac{\partial}{\partial t}\; y^{(t)}(\vp ) = v^{(t)}(y^{(t)}(\vp) ) \; .
\ee
The integral form of the latter can be presented as the evolution integral
\be
\label{37}
\int_{y^{(L)}}^{y^{(L)*}} \; \frac{d\vp}{v^{(L)}(\vp,\qq )} =
\frac{1}{L} \; .
\ee
If the parameter $q$ is unknown, it must be determined from a simultaneous
treatment of the $ \beta $-function (\ref{20}). In this case we determine
a sequence of optimal parameters $\qq =\qq ^{(L)*}$, leading to the
self-similar approximants
\be
\label{38}
f^{(L)*}(g_B) \equiv F^{(L)*}(g_B,s^{(L)}(g_B),\qq ^{(L)*}) \; .
\ee
For the purpose of determining critical exponents we are only interested
in $f(g_B)$ at $g_B\ra\infty$ and  go to the limit of Eq.~(\ref{38})
yielding
\be
\label{39}
f^{(L)*} \equiv \lim_{g_B\ra\infty} f^{(L)*}(g_B) \; .
\ee
The self-similar approximant (\ref{39}) replaces the previous optimized
approximant (\ref{19}) of field-theoretic variational perturbation theory.

From the definition of the pesudovelocity (\ref{35}) it follows that for
the calculation of the $L$-order self-similar approximant $f^{(L)*}$, we
need to know $L+1$ orders of the expansion in (\ref{1}). If $L$ is the last
available order, we shall use as an $L+1$st approximation the average of
the previous ones:
\be
\label{40}
f^{(L+1)*} = \frac{1}{2}\; (f^{(L-1)*} + f^{(L)*} ) \; .
\ee
This approximation is expected to be reliable if the approximants tend
to the limit $L\rightarrow\infty$ in an alternating fashion, once from
above and once from below. This is not a priori ensured, but happens in
many examples (but there exist also counter-examples). In the series for
the critical exponents to be treated here this seems to be true. Thus
Eq. (\ref{40}) will be used to get  the highest approximant. More refined
mathematical foundation for the usage of Eq. (\ref{39}) is given in Refs.
\cite{b22,b23}.

\section{IV. Combining Self-Similar and Variational Theories}

Let us now be explicit and improve the convergence of the sequence
$\{ f^{(L)\rm \,opt}\}$ of variational perturbation theory derived in
(\ref{16})--(\ref{19}) by self-similar approximation theory to obtain
a new sequence $\{ f^{(L)*}\}$. The improvement is most drastic at the
initial stages of the procedure, when $L\leq 3$, so that we shall restrict
ourselves to these low orders. An additional advantage is that all
formulas up to third order can be derived analytically.

Recall that, in contrast to Section II, we do not consider from the
beginning the limit of $g_B\ra\infty$, but retain the full $g_B$-dependence
of the functions (\ref{7}):
\begin{eqnarray}
F^{(1)}(g_B,s,\qq ) &=& f_0 + f_1\; s \; , \label{41}\\ [.5em]
F^{(2)}(g_B,s,\qq ) &=& F^{(1)}(g_B,s,\qq ) +\qq  f_1\sgm\; s + f_2\; s^2 \; ,\\
\label{42}
F^{(3)}(g_B,s,\qq ) &=& F^{(2)}(g_B,s,\qq ) +\frac{1}{2}\; \qq (1+\qq )
f_1\sgm^2\; s + 2\qq  f_2\sgm \; s^2 + f_3\; s^3 \; ,
\label{43}
\end{eqnarray}
which reduce to Eqs. (\ref{8})--(\ref{10}) for $g_B\ra\infty$ since then
$\sgm\ra 1$. For arbitrary $g_B$ we must optimize $F^{(L)}$ in $s$. Since
$\sigma $ depends on $s$ via the relation (\ref{4}), we may look for the
extremum in the two $s$ and $\sigma$ while satisfying the condition
\be
\label{44}
\frac{d\sgm}{d s} = \frac{\sgm-1}{\qq  s} \; .
\ee
If this is done with the function $F^{(L)}$, we obtain  an optimal function
$s^{(L)}(g_B)$. From this we calculate  the approximant
\be
\label{45}
F^{(L)\,{\rm opt}}(g_B,\qq ) \equiv F^{(L)}(g_B,s^{(L)}(g_B),\qq ) \; .
\ee
To lowest order $L=1$, an optimal function  dos usually not exist. In this
case we shall use the next higher existing $s^{(2)}(g_B)$ to define the
lowest approximant. In principle we could, of course, form an entire
off-diagonal matrix of variational functions
\be
\label{46}
F^{(L,L')}(g_B,\qq ) \equiv F^{(L)}(g_B,s^{(L')}(g_B),\qq ) \; .
\ee
of which the functions (\ref{16}) in  Section II are only diagonal elements
\be
\label{47}
F^{(L)\rm \,opt}(\infty,\qq ) = F^{(L,L)}(\infty,\qq ) \; .
\ee

The optimal function $s^{(2)}(g_B)$  is determined by the extremality
condition (\ref{27}), which amounts to the equation
\be
\label{48}
(1+\qq )f_1\sgm(g_B) + 2f_2\; s = 0 \; .
\ee
From this we obtain the variational expression
\be
\label{49}
F^{(2)\,{\rm opt}}(g_B,\qq )
=F^{(2,2)}(g_B,\qq )
= f_0 + f_1\; s^{(2)} +
\frac{1-\qq }{1+\qq }\; f_2\; s^{(2)\,2} \; .
\ee
In addition, we determine a lowest approximant with a first-order trial
function $s^{(1)}(g_B)=s^{(2)}(g_B)$,
\begin{eqnarray}
\label{50}
F^{(1)\,{\rm opt}}(g_B,\qq )
=F^{(1,2)}(g_B,\qq ) = f_0 + f_1\; s^{(2)} \; .
\end{eqnarray}

The third-order trial function $s^{(3)}{}(g_B)$ is given either by the
extremum-point condition (\ref{27}), which yields
\be
\label{51}
\frac{1}{2}\; (1+\qq )(2+\qq ) f_1\sgm^2 + 2(1+2\qq )f_2\sgm s^{(3)}{} +
3f_3 s^{(3)}{}^2 = 0 \; ,
\ee
or,
when Eq. (\ref{51}) has no real solution, by the turning point condition
(\ref{28}), by solving
\be
\label{52}
(1+\qq )(2+\qq )f_1\sgm(\sgm-1) + 2(1+2\qq ) f_2\; s^{(3)}{}
(\sgm-1+\qq \sgm) + 6f_3\qq  s^{(3)}{}^2 = 0 \; .
\ee

We now construct the approximation cascade following the rules of Section
III. Accordingly, we set $y^{(1)}=F^{(1)}$, $y^{(2)}=F^{(2)}$, and
$y^{(L)}=F^{(L)}$ for $L\geq 2$. Therefore the reonomic-constraint (\ref{29})
reads $f_0+f_1s=\vp$, which defines the reonomic function $g_B^{(1)}(\vp)$
through the relation
\be
\label{53}
s^{(2)}(g_B(\vp)) = \frac{\vp -f_0}{f_1} \; .
\ee
The first-order psudovelocity as defined in Eq. (\ref{35}) is
$v^{(1)}=F^{(2,1)}-F^{(1,1)}$, which reads explicitly
\be
\label{54}
v^{(1)}(\vp,\qq ) = A_1(\vp-f_0)^2 \; ,
\ee
with
\be
\label{55}
A_1 \equiv \frac{(1-\qq )f_2}{(1+\qq )f_1^2} \; .
\ee
Using the evolution integral (\ref{37}), we find from self-similar approximant
(\ref{33}) the expression
\be
\label{56}
F^{(1)*}(g_B,\qq ) = f_0 +
\frac{F^{(2,2)}(g_B,\qq )-f_0}{1-A_1[F^{(2,2)}(g_B,\qq )-f_0]} \; ,
\ee
where $F^{(2,2)}$ of Eq. (\ref{45}) is given by Eq. (\ref{49}). In the
strong-coupling limit $g_B\ra\infty$, this reduces to
\be
\label{57}
F^{(1)*}(\infty,\qq ) = f_0 - \; \frac{(1+\qq )^2 f_1^2}{(5-\qq )^2f_2} \; .
\ee

Let us now determine the value of the parameter $\qq $, following the
procedure described at the end of Section II, but with the difference that
now we construct the self-similar approximants for the expansion (\ref{21})
of the $\beta$-function, which will be denoted by $B^{(L)*}(g_B,\qq )$.
These have the same form as $F^{(L)*}(g_B,\qq )$, except that the expansion
coefficients $f_n$ are replaced by $\beta_n$. The  parameters $\qq ^{(L)*}$
are defined by the boundary condition
\be
\label{58}
B^{(L)*}(\infty,\qq ^{(L)*}) = 0 \; .
\ee
To first order, the result is
\be
\label{59}
\qq ^{(1)*} =
\frac{\sqrt{\beta_0 \beta_2(4\beta_1^2 + 5\beta_0\beta_2)} -
\beta_1^2}{\beta_1^2 + \beta_0 \beta_2} \; .
\ee
Substituting this into Eq.\ (\ref{56}), we obtain
$f^{(1)*}(g_B)=F^{(1)*}(g_B,\qq ^{(1)*})$, in agreement with Eq.(\ref{34}).
The final result is given by Eq.\ (\ref{39}),  that is, by the value
$f^{(1)*}=f^{(1)*}(\infty)$.

The second-order  velocity (\ref{35}) is $v^{(2)}=F^{(3,2)}-F^{(2,2)}$, where
\be
\label{60}
F^{(3,2)}(g_B,\qq ) = F^{(2,2)}(g_B,\qq ) +
\frac{(1+\qq )f_1f_3-2\qq  f_2^2}{(1+\qq )f_1}\; s^{(2)\,3}\; ,
\ee
which is valid for any $g_B$. In particular, in the limit $g_B\ra\infty$, where
\be
\label{61}
F^{(3,2)}(\infty,\qq ) = f_0 - (1+\qq )^3 \; \frac{f_1^2}{4f_2}
\left ( \frac{f_1 f_3}{2f_2^2} \; - \; \frac{1-\qq }{1+\qq }
\right ) \; .
\ee
The explicit form of the velocity $v^{(2)}$ is now
\be
\label{62}
v^{(2)}(\vp,\qq ) = A_2 (\vp-f_0)^3 \; ,
\ee
where
\be
\label{63}
A_2 \equiv \frac{(1+\qq )f_1 f_3 -2\qq  f_2^2}{(1+\qq ) f_1^4} \; .
\ee
From the evolution integral (\ref{37}) we find here
\be
\label{64}
F^{(2)*}(g_B,\qq ) = f_0 +
\frac{F^{(2,2)}(g_B,\qq )-f_0}{\sqrt{1-A_2[F^{(2,2)}(g_B,\qq )-f_0]^2}} \; .
\ee
The value $\qq ^{(2)*}$ follows from the strong-coupling condition (\ref{58}),
which leads to the equation:
\be
\label{65}
(1+\qq ^{(2)*})^4 (\beta_0^2 \beta_1 \beta_3 - 2\beta_0^2\beta_2^2 + \beta_1^4)
+ 2(1+\qq ^{(2)*})^3 \beta_0^2 \beta_2^2 - 16 \beta_0^2 \beta_2^2 = 0 \; .
\ee
Inserting the appropriate solution $\qq ^{(2)*}$ into Eq. (\ref{64}), we obtain
$f^{(2)*}(g_B)=F^{(2)*}(g_B,\qq ^{(2)*})$. And the limiting value (\ref{39})
is $f^{(2)*}=f^{(2)*}(\infty)$. If only three orders of the expansion (\ref{1})
are available, then $f^{(3)*}$ is defined by Eq. (\ref{40}), as explained at
the end of Section III.

\section{V. Application to Critical Exponents}

The above theory will now be applied to evaluate the divergent perturbation
expansions of the critical exponents of $O(N)$-symmetric $\vp^4$-field
in $4-\ep$ dimensions. The expansions are power series in the bare coupling
parameter $g_B/\mu^\epsilon$, where $\mu$ is some mass parameter to make
$g_B/\mu^ \epsilon $ dimensionless. They can be found up to six loops in the
textbook \cite{b2}. In this field-theoretic context, the parameter $\qq$ in
the transformation (\ref{3}) is directly related to the Wegner exponent
$\omega$ \cite{b24} which characterizes the strong-coupling behaviour of
the renormalized coupling
\be
\label{66}
g(g_B) \simeq g(\infty) - \;{\rm const}\times
 \frac{\mu^ \omega }{g_B^{\om/\ep}} \qquad
(g_B/\mu^ \epsilon \ra\infty) \; .
\ee
The relation is
\be
\label{67} \!\!\!\!\!\!\!\!\!\!\!\!\!\!\!\! \!\!\!\!\!\!\!
\om = \ep/\qq  \; .
\ee
In the sequel, we shall set $\mu=1$. Starting point is the expansion
\cite{b2} for the renormalized coupling constant which we shall limit to
$g_B^4$, for simplicity;
\be
\label{68}       \!\!\!\!\!\!\!\!\!\!
g_R(g_B) \simeq g_B + c_2 g_B^2 + c_3 g_B^3 + c_4 g_B^4 \; ,
\ee
where $g_B\ra 0$ and the coefficients are
\begin{eqnarray}
\!\!\!\!\!\!\!\!\!\!\!\!\!\!\!\!\!\!\!\!\!\!\!\!
\!\!\!\!\!\!\!\!\!\!\!\!\!\!\!\!c_0 = 0 \; , \qquad c_1 = 1 \; , \qquad
c_2 = -\; \frac{N+8}{3\ep} \; , \qquad
c_3 = \frac{(N+8)^2}{9\ep^2} + \frac{3N+14}{6\ep} \; , \nonumber
\end{eqnarray}
\begin{eqnarray}
\label{69}
c_4 = -\; \frac{(N+8)^3}{27\ep^3} \; - \;
\frac{4(N+8)(3N+14)}{27\ep^2} \; -
\; \frac{33N^2+922N +2960 + 24(27N+88)\zeta(3)}{648\ep} \; ,
\end{eqnarray}
with $\zeta(z)$ being the Riemann zeta function. The logarithmic derivative
of (\ref{68}) yields the $\beta$-function
\be
\label{70}  \!\!\!\!\!\!\!\!\!\!\!\!
\bt^{(3)}(g_B) =
\frac{d\log g(g_B)}{d\log g_B}= 1 + \beta_1 g_B + \beta_2 g_B^2 +
\beta_3 g_B^3 \; ,
\ee
with the coefficients
\be
\label{71}
\beta_1 = c_2 \; , \qquad \beta_2 = 2c_3 - c_2^2 \; , \qquad
\beta_3 = c_2^3 - 3c_2 c_3 + 3c_4 \; .
\ee
In second-order variational perturbation theory, we have $\qq^{(2)}$ given
by Eq. (\ref{26}), which yields
\be
\label{72}
\qq^{(2)} = 2\sqrt{1+p\ep} - 1 \; ,
\ee
where the notation
\be
\label{73}
p \equiv \frac{3(3N+14)}{(N+8)^2}
\ee
is used. The first-order self-similar approximant $\qq^{(1)*}$ is defined in
Eq. (\ref{59}), resulting in
\be
\label{74}
\qq^{(1)*} = \frac{\sqrt{(1+p\ep)(9+5p\ep)}-1}{2+p\ep} \; .
\ee
The corresponding Wegner exponents are
\be
\label{75}
\omega^{(2)} = \frac{\ep}{2\sqrt{1+p\ep}-1} \; ,
\ee
in the variational perturbation theory, and
\be
\label{76}
\omega^{(1)*} = \frac{(2+p\ep)\ep}{\sqrt{(1+p\ep)(9+5p\ep)} - 1} \; ,
\ee
in the self-similar approximation theory.

The second-order optimized approximant
\be
\label{77}
G^{(2)\,\rm opt}(\infty,\qq ) = \frac{3(1+\qq )^2\ep}{4(N+8)} \; ,
\ee
corresponding to the renormalized coupling (\ref{68}), with the parameter
(\ref{72}), becomes
\be
\label{78}
g^{(2)\,\rm opt} = \frac{3(\ep+p\ep^2)}{N+8} \; ,
\ee
according to Eq. (\ref{19}). The first-order self-similar approximant
\be
\label{79}
G^{(1)*}(\infty,\qq ) =
\frac{3(1+\qq )^2\ep}{(5-\qq ^2)(N+8)} \; ,
\ee
with $\qq^{(1)*}$ from Eq. (\ref{74}), results in
$g_1^*=G_1^*(\infty,\qq^{(1)*})$, as in Eq.\ (\ref{39}). It turns out that,
because of the equality
\be
\label{80}
\frac{(1+\qq^{(1)*})^2}{5-(\qq^{(1)*})^2} = 1 + p\ep \; ,
\ee
the values $g^{(1)*}$ and $g^{(2)\,\rm opt}$ coincide. However, $g^{(2)*}$,
following from Eq. (\ref{64}), is different from $g^{(3)\,\rm opt}$.

We now turn to the perturbation expansions of the critical exponents $\nu$
and $\gm$. Other exponents need not be treated since they can be found from
the above using well-known scaling relations (see e.g. \cite{b2,b25,b26}).
We begin with $\nu^{-1}$ for which we use the expansion \cite{b2} up to
$g_B^3$
\be\!\!\!\!\!\!\!\!
\label{81}
\nu^{-1}{}^{(3)} = f_0 + f_1 g_B + f_2 g_B^2 + f_3 g_B^3 \; ,
\ee
with the coefficients
\be
\label{82}
f_0 = 2 \; , \qquad f_1 = -\; \frac{N+2}{3} \; , \qquad
f_2 = \frac{N+2}{9} \left ( \frac{N+8}{\ep} + \frac{5}{2}\right ) \; ,
\ee
\be
\label{83}
f_3 = -\; \frac{N+2}{108} \left [ \frac{4(N+8)^2}{\ep^2} +
\frac{2(19N+122)}{\ep} + 3(5N+37) \right ] \; .
\ee
Following the above procedure, we get the optimized strong-coupling value
\be
\label{84}
\nu^{-1}{}^{(2)*} =
2 - (1+\qq )^2\; \frac{(N+2)\ep}{2(2N+16+5\ep)} \; ,
\ee
which, for $\qq^{(2)}$ from Eq.\ (\ref{72}), gives the variational perturbation
result
$\nu^{-1}{}^{(2)\,\rm opt}=\nu^{-1}{}^{(2)\,\rm opt}(\infty,\qq^{(2)})$.
Its self-similar improvement reads
\be
\label{85}
\nu^{-1}{}^{(1)*}(\infty,\qq ) = 2 - \;
\frac{2(1+\qq )^2(N+2)\ep}{(5-\qq ^2)(2N+16+5\ep)} \; .
\ee
After inserting $\qq^{(1)*}$ from Eq.\ (\ref{74}), we obtain
\be
\label{86}
\nu^{-1}{}^{(1)*}
 = 2 - \; \frac{2(N+2)(1+p\ep)\ep}{2(N+8)+5\ep} \; .
\ee
Again, it turns out that $\nu^{-1}{}^{(1)*}=\nu^{-1}{}^{(2)\,\rm opt}$,
but $\nu^{-1}{}^{(2)*}\neq \nu^{(3)\,\rm opt}$, with $\nu^{-1}{}^{(2)*}$
defined by Eq. (\ref{49}).

Finally, we resum the perturbation expansion for the critical exponent
$\gamma=\nu(2- \eta)$, which reads in the form \cite{b2}:
\be
\label{87}
\gm(g_B) \simeq f_0 + f_1 g_B + f_2 g_B^2 + f_3 g_B^3 \; ,
\ee
with the coefficients
\be
\label{88}
f_0=1\; , \qquad f_1 = \frac{N+2}{6} \; , \qquad
f_2 = -\; \frac{N+2}{36}\left [ \frac{2(N+8)}{\ep} + 4 - N\right ] \; ,
\ee
\be
\label{89}
f_3 = \frac{N+2}{432} \left [ \frac{8(N+8)^2}{\ep^2} +
\frac{4(106+N-2N^2)}{\ep} + 194 + N(2N+ 17) \right ] \; .
\ee
Here we find the optimized approximant
\be
\label{90}
\gm^{(2)\,\rm opt}(\infty,\qq ) = 1 +
\frac{(1+\qq )^2(N+2)\ep}{4[2(N+8)+(4-N)\ep]}
\ee
and the self-similar approximant
\be
\label{91}
\gm^{(1)* }= 1 + \frac{(N+2)(1+p\ep)\ep}{2(N+8)+(4-N)\ep} \; .
\ee
Again $\gm^{(1)* }$ coincides with the variational perturbation result
$\gm^{(2)\,\rm opt}=\gm^{(2)\,\rm opt}(\infty,\qq^{(2)})$, whereas
$\gm^{(2)*}$ does not equal $\gm^{(3)\,\rm opt}$.

By construction, the self-similar approximants $f^{(L)*}$ obtained from
the evolution integral (\ref{30}) possess the same $\ep$-expansion, up to
the given order $L$, as the optimized approximant $f^{(L)\rm \,opt}$ of
variational perturbation theory. This is evident, from expressions (\ref{56})
and (\ref{64}). For the variational perturbation results, on the other hand,
it was shown in Ref.~\cite{b4,b5,b6,b19} that all expansions in powers of
$\epsilon$ coincide with the expansions derived in the  renormalization-group
approach to critical phenomena \cite{b2}. As a consequence, also the presently
derived self-similar approximants $f^{(L)*}$ possess the exact $\ep$-expansions.
This can easily  be verified by an explicit calculation.

It is  interesting to compare the $1/N$-expansions with the self-similar
approximants for larger $N$. These expansions for the critical exponents
$\om$, $\nu$, $\gm$, and $\eta$ are presented in Fig. 1, where they are compared
with our self-similar approximants as well as with the results of the sixth-order
variational perturbation theory \cite{b2}, and with those of Pad\'e-Borel
resummations \cite{b27,b28}. Our third-order results have the same accuracy as
those of sixth or higher orders, obtained by other resummation techniques. In
the limit $N\ra\infty$, our exponents coincide with the known exact values
\be
\label{92}
\al= \frac{D-4}{D-2} \; , \qquad \bt= \frac{1}{2}\; , \qquad
\gm= \frac{2}{D-2} \; , \qquad \dlt=\frac{D+2}{D-2} \; ,
\ee
\be
\label{93}
\nu= \frac{1}{D-2} \; , \qquad \eta= 0 \; , \qquad \om=4-D \; ,
\ee
where $D$ is dimensionality.

Critical exponents for finite $N$ have been calculated by Pad\'e-Borel
resummation methods based on six- and seven-loop expansions in $D=3$
dimensions \cite{b29,b30,b31,b32} or in five-loop expansions in $D=4-\ep$
dimensions \cite{b33,b34,b35,b36}. Different variants of the optimized
perturbation theory \cite{b20} have been used \cite{b37,b38,b39,b40}.
Self-similar exponential approximants were given in \cite{b41}. Computer
simulations, based on the Monte Carlo lattice studies were presented in
\cite{b42}. The available results have been reviewed in Refs. \cite{b2,b43,b44}.

A list of our results from the third-order self-similar improvements of
variational perturbation theory for the critical exponents is given in Table 1.
The exponents $\nu$, $\gm$, and $\om$ are calculated directly from their series,
as is explained in the text. The other listed exponents are obtained from the
scaling relations
\be
\label{94}
\al = 2-\nu D \; , \qquad \bt =\frac{\nu}{2}\; (D-2+\eta) \; , \qquad
\gm=\nu(2-\eta) \; .
\ee
The error bars are defined by the difference between $f^{(L)*}$ and
$f^{(L-1)*}$. Our results are compared with those of the field-theoretic
variational perturbation theory based on six-loop (for $N>3$) and
seven-loop (for $N=0,1,2,3$) expansions calculated in three dimensions
in Refs. \cite{b3,b4,b5} and listed in the book \cite{b2}. We also
show the exponents recently obtained by Pad\'e-Borel resummation \cite{b28},
as well as earlier results \cite{b29,b30,b31,b32,b33,b34,b35,b36}, based
on six-loop expansions in $D=3$ dimensions \cite{b29,b30,b31,b32}
and on five-loop expansions in $D=4-\ep$ dimensions \cite{b33,b34,b35,b36}.
The numbers in parentheses indicate the highest calculated approximation
(seventh order for $n=0,1,2,3$ and sixth order for $N>3$), from which the
effective extrapolations to infinite order were obtained as described in
the book \cite{b2}. Comparing the results, we see that our third-order
self-similar approximants yield practically the same values for the
critical exponents as the values derived by other resummation techniques
of sixth or seventh order. In the limiting cases of $N=-2$ and $N=\infty$,
our results coincide with the known exact values of the critical exponents.

The critical exponents obtained by our new method from three-loop
perturbation expansions are all in good agreement with all experiments.
Unfortunately, most of them are not sufficiently accurate to distinguish
between different theoretical approaches. The most accurately known
experiment is the measurement of the specific heat of liquid helium with
nanoKelvin temperature resolution near the lambda point, which were
performed in a satellite orbiting around the Earth  \cite{b45,b46}. The
specific-heat exponent  initially extracted from the data in \cite{b45}
was $\al=-0.01056\pm0.0004$. This differed slightly from the result of
seven-loop variational perturbation theory obtained from three-dimensional
$\phi^4$-theory which yielded $\al=-0.0129\pm 0.0006$ \cite{b5}. However,
a recently performed re-analysis of the data in Ref.~\cite{b46} found
$\al=-0.0127\pm 0.0003$, thus confirming with great precision the
theoretical result of Ref.~\cite{b5}. Our present result $\al=-0.0124$
obtained in third-order self-similar-improved variational perturbation
theory is again in perfect agreement with the latest experimental result.
This is quite remarkable since the five-loop calculations in $4-\ep$
dimensions gave the value $\al=-0.013$ \cite{b41}. This illustrates the
acceleration of the convergence of variational perturbation theory by the
self-similar improvement developed in this
paper.

\vskip 5mm

In conclusion, we have developed a novel method for resuming divergent
perturbation expansions. It combines field-theoretical variational
perturbation theory with the self-similar approximation theory, and
accelerates greatly  the convergence of either method alone. The
acceleration is especially useful if only  low-order expansions are
available due to the complexity of the problem. Up to third order, all
results are found analytically. The method was illustrated by calculating
the critical exponents whose third-order approximants are found to be
practically the same as the sixth- or seventh-order approximants of
other resummation techniques. The specific-heat critical exponent $\al$
for $N=2$ is found to be in perfect agreement with the most accurately
measured experimental value of $\al$ for superfluid helium.

\vskip 5mm

{\bf Acknowledgement}

\vskip 2mm

One of the authors (V.I.Y.) is grateful to A. Pelster and E.P. Yukalova
for discussions. Financial support from the German Academic Exchange
Service (DAAD) is appreciated.

\begin{table}[tbhp]
\caption[
Third-order critical exponents of self-similar variational perturbation
results of this paper
]{
Third-order critical exponents of self-similar variational
perturbation results of this paper obtained from three-loop expansions
in $4-\epsilon$ dimensions. Results are compared with the six-loop
(for $N>3$) exponents and seven-loop exponents (for $N=0,1,2,3$) calculated
in three dimensions in Refs. \cite{b3,b4,b5} and listed in the textbook
\cite{b2}. We also show the exponents obtained by Pad\'e-Borel resummation
in Ref. \cite{b28}, as well as earlier results (all cited in Notes and
References). They refer to six-loop expansions in $D=3$ dimensions
\cite{b29,b30,b31,b32}, or to five-loop expansions in $\epsilon=4-D$
\cite{b34,b35}. The numbers in parentheses indicate the highest calculated
approximation (seventh order for $N=0,1,2,3$ and sixth order for $N>3$)
from which the final results were obtained by extrapolation to infinite
order. The critical couplings $g_c$ are different for calculations in
$4-\epsilon$ and three dimensions due to different normalizations.
}

\newpage
\label{TableI}\end{table}
\tabcolsep1.78mm
\scriptsize{\begin{tabular}{l|llllll|ll}
\hline \hline
~$N$ & $g_c$ & $ \gamma $ & $ \eta $ & $ \nu$ & $  \alpha $ & $  \beta $& $\omega$ ($\omega_6$)\\
\hline
 -2& 0.758$\pm$0.037  & 1    & 0 & 1/2  &2-$D$/2 &($D$-2)/4  & 0.831$\pm$ 0.077   &   \\[0mm]
\hline
 ~0& 0.578$\pm$0.021  & 1.161$\pm$ 0.004    & 0.028$\pm$0.005 & 0.588$\pm$0.001  & 0.235$\pm$ 0.001  & 0.311$\pm$0.001  & 0.812$\pm$ 0.055   &   \\[0mm]
 &    & 1.161(1.159)   & 0.0311$\pm$0.001 & 0.5886(0.5864) &0.234 &  &0.810(0.773)&\protect\cite{b5}                   \\[0mm]
 &  & 1.168(1.159)   & 0.025(0.0206) & 0.592(0.586) & &  &0.810(0.7737)    &\protect\cite{b2}     \\[0mm]
                     & 1.402$^{\rm }$   & 1.160$^{\rm }$& 0.034$^{\rm }$& 0.589$^{\rm }$& 0.231$^{\rm }$  & 0.305$^{\rm }$                                                                              &&\protect\cite{b27,b28}  \\
                     & 1.421$\pm0.004^{\rm }$   & 1.161$\pm0.003^{\rm }$ & 0.026$\pm0.026^{\rm }$ & 0.588$\pm0.001^{\rm }$ & 0.236$\pm0.004^{\rm }$ & 0.302$\pm0.004^{\rm }$ & $0.794\pm0.06^{\rm }$     &\protect\cite{b29}      \\
                     &$ 1.421\pm 0.008^{\rm }$ &$ 1.1615\pm 0.002^{\rm }$ & $0.027\pm 0.004^{\rm }$ & 0.5880$\pm 0.0015^{\rm }$ &      &  0.3020$\pm 0.0015^{\rm }$& $0.80\pm 0.04^{\rm }$                &\protect\cite{b30,b31,b32}      \\
                     & &$ 1.160\pm 0.004^{\rm }$ & $0.031\pm 0.003^{\rm }$ & 0.5885$\pm 0.0025^{\rm }$ &      &  0.3025$\pm 0.0025^{\rm }$& $0.82\pm 0.04^{\rm }$                                          &\protect\cite{b36}     \\
\hline
~1& 0.510 $\pm$ & 1.238$\pm$0.004  & 0.037$\pm$0.009   & 0.630$\pm$0.008 & 0.109$\pm$0.012  & 0.327$\pm$0.004 & 0.808$\pm$(0.046& \\[0mm]
&  & 1.241(1.236)  &0.0347$\pm0.001$   & 0.6310(0.6270) &0.107 &  &0.805(0.772)&\protect\cite{b5}\\[0mm]
&  & 1.241(1.235)  &0.030(0.0254)   & 0.630(0.627) & &  &0.805(0.7724)&\protect\cite{b2}\\[0mm]
& 1.419$^{\rm }$& 1.239$^{\rm }$& 0.038$^{\rm }$ & 0.631$^{\rm }$ & 0.107$^{\rm }$& 0.327$^{\rm }$  &0.781$^{\rm }$                                                                         &\protect\cite{b27,b28}  \\
  & 1.416$\pm 0.0015^{\rm }$ & 1.241$\pm 0.004^{\rm }$ & 0.031$\pm 0.011^{\rm }$ & 0.630$\pm 0.002^{\rm }$& $0.110\!\pm\!0.008^{\rm }$&$ 0.324\pm0.06^{\rm }$& $0.788\!\pm\!0.003^{\rm }$   &\protect\cite{b29}      \\
  & 1.416$\pm$0.004$^{\rm }$  & 1.2410$\pm$0.0020$^{\rm }$ &  0.031$\pm$0.004$^{\rm }$ & 0.6300$\pm$0.0015$^{\rm }$ &   & 0.3250$\pm$0.0015$^{\rm }$ &0.79$\pm$0.03$^{\rm }$                &\protect\cite{b30,b31,b32}      \\
  & &  & $0.035\pm 0.002^{\rm }$ & 0.628$\pm 0.001^{\rm }$ &      & & $0.80\pm 0.02^{\rm }$                                                                                                 &\protect\cite{b33,b34,b35}  \\
  & &$ 1.1239\pm 0.004^{\rm }$ & $0.037\pm 0.003^{\rm }$ & 0.6305$\pm 0.0025^{\rm }$ &      &  0.3265$\pm 0.0025^{\rm }$& $0.81\pm 0.04^{\rm }$                                             &\protect\cite{b36}     \\
\hline
~2 & 0.454$\pm$0.012  & 1.310$\pm$0.019 & 0.045$\pm$0.012  & 0.671$\pm$0.018 & -0.0124$\pm$0.0270   & 0.343$\pm$0.009 &0.807$\pm$0.038&    \\[0mm]
 &  & 1.318(1.306) & $0.0356\pm0.001$  & 0.6713(0.6652) &-0.0129  & &0.800(0.772)& \protect\cite{b5}                    \\[0mm]
 &  & 1.318(1.306) & 0.032(0.0278)  & 0.670(0.665) &  & &0.800(0.7731)&\protect\cite{b2}   \\[0mm]
 & 1.408$^{\rm } $& 1.315$^{\rm }$ & 0.039$^{\rm } $& 0.670$^{\rm }$ & -0.010$^{\rm }$& 0.348$^{\rm }$&0.780$^{\rm }$                                                                      &\protect\cite{b27,b28} \\
   & 1.406$\pm$0.005$^{\rm }$ & 1.316$\pm$0.009$^{\rm }$ & 0.032$\pm$0.015$^{\rm }$ & 0.669$\pm$0.003$^{\rm }$ & -0.007$\pm$0.009$^{\rm }$ & 0.346$\pm$0.009$^{\rm }$&$0.78\pm0.01^{\rm }$ &\protect\cite{b29}     \\
  & 1.406$\pm$0.004$^{\rm }$ & 1.3160$\pm$0.0025$^{\rm }$ & 0.033$\pm$0.004$^{\rm }$ & 0.6690$\pm$0.0020$^{\rm }$   &     & 0.3455$\pm$0.002$^{\rm }$ &0.78$\pm$0.025$^{\rm }$             &\protect\cite{b30,b31,b32}     \\
  & &  & $0.037\pm 0.002^{\rm }$ & 0.665$\pm 0.001^{\rm }$ &      & &$0.79\pm 0.02^{\rm }$                                                                                                 &\protect\cite{b33,b34,b35}  \\
  & &$ 1.315\pm 0.007^{\rm }$ & $0.040\pm 0.003^{\rm }$ & 0.671$\pm 0.005^{\rm }$ &      &  0.3485$\pm 0.0035^{\rm }$& $0.80\pm 0.04^{\rm }$                                               &\protect\cite{b36}    \\
\hline
 ~3& 0.407$\pm$0.010  &1.378$\pm$0.037  & 0.052$\pm$0.015  & 0.709$\pm$0.030 & -0.126$\pm$0.045  & 0.359$\pm$0.013 & 0.807$\pm$0.031 &\\[0mm]
 &  &1.390(1.374)  &$0.0350\pm0.0005$  & 0.7072(0.7004) &-0.122  & &0.797(0.776)&\protect\cite{b5}     \\[0mm]
  &  &1.387(1.372)  &0.032(0.0288)  & 0.705(0.700) &  & &0.797(0.7758) &\protect\cite{b2}\\[0mm]
 & 1.392$^{\rm }$ & 1.386$^{\rm }$ & 0.038$^{\rm }$ & 0.706$^{\rm }$ & -0.117$^{\rm }$ & 0.366$^{\rm }$&0.780$^{\rm }$                                                                     &\protect\cite{b27,b28} \\
  & 1.392$\pm$0.009$^{\rm }$ & 1.390$\pm$0.01$^{\rm }$ &  0.031$\pm0.022^{\rm }$ &  0.705$\pm$0.005$^{\rm }$ & -0.115$\pm0.015^{\rm }$ & 0.362$^{\rm }$ &$0.78\pm0.02^{\rm }$              &\protect\cite{b29}     \\
  & 1.391 $\pm$0.004$^{\rm }$ & 1.386$\pm$0.004$^{\rm }$ & 0.033$\pm$0.004$^{\rm }$ & 0.705$\pm$0.003$^{\rm }$ &    &    0.3645$\pm$0.0025$^{\rm }$&$0.78\pm0.02^{\rm } $                  &\protect\cite{b30,b31,b32}     \\
  & &  & $0.037\pm 0.002^{\rm }$ & 0.79$\pm 0.02^{\rm }$ &      & & $0.79\pm 0.02^{\rm }$                                                                                                  &\protect\cite{b33,b34,b35}  \\
  & &$ 1.390\pm 0.010^{\rm }$ & $0.040\pm 0.003^{\rm }$ & 0.710$\pm 0.007^{\rm }$ &      &  0.368$\pm 0.004^{\rm }$& $0.79\pm 0.04^{\rm }$                                                 &\protect\cite{b36}    \\
\hline
 ~4 & 0.368$\pm$0.008 & 1.442$\pm$0.056 & 0.057$\pm$0.018 & 0.744$\pm$0.043 & -0.232$\pm$0.064 & 0.374$\pm$0.018  & 0.809$\pm$0.026& \\
 &           & 1.451(1.433) & 0.031(0.0289) & 0.737(0.732)&& & 0.795(0.780)&\protect\cite{b2} \\
 & 1.375$^{\rm }$ & 1.449$^{\rm }$ & 0.036$^{\rm }$& 0.738$^{\rm }$ & -0.213$^{\rm }$& 0.382$^{\rm }$& 0.783$^{\rm }$
&\protect\cite{b27,b28}\\
\hline
  ~5 & 0.335$\pm$0.007  & 1.501$\pm$0.076 & 0.060$\pm$0.019 & 0.776$\pm$0.055  & -0.328$\pm$0.082  & 0.388$\pm$0.022 & 0.812$\pm$0.022 &  \\
  &  &1.511(1.487)  & 0.0295(0.0283)& 0.767(0.760) &  & &0.795(0.785)&\protect\cite{b2}  \\
 & 1.357$^{\rm }$& 1.506$^{\rm }$& 0.034$^{\rm }$ & 0.766$^{\rm }$ & -0.297$^{\rm }$& 0.396$^{\rm }$&0.788$^{\rm }$&\protect\cite{b27,b28}\\
\hline
 ~6 & 0.306$\pm$0.006   & 1.554$\pm$ 0.095 & 0.062$\pm$0.020) & 0.804$\pm$0.066 & -0.414$\pm$0.099  & 0.399$\pm$0.025 &0.814$\pm$0.019 (0.792)& \\
  &  &1.558(1.535) & 0.0276(0.0273) & 0.790(0.785) &  &  &0.797(0.792)&\protect\cite{b2} \\
 & 1.339$^{\rm }$ & 1.556$^{\rm }$ & 0.031$^{\rm }$ & 0.790$^{\rm }$& -0.370$^{\rm }$& 0.407$^{\rm }$ &0.793$^{\rm }$&\protect\cite{b27,b28}\\
\hline
~7 & 0.282$\pm$0.005 & 1.601$\pm$0.112 & 0.062$\pm$0.020 & 0.829$\pm$0.075 & -0.489$\pm$0.0113  & 0.409$\pm$0.028      & 0.818$\pm$ 0.016&  \\
 & & 1.599(1.577) & 0.0262(0.0260) & 0.810(0.807)&  &  &0.802(0.800)&\protect\cite{b2}  \\
 & 1.321$^{\rm }$ & 1.599$^{\rm }$& 0.029$^{\rm }$& 0.811$^{\rm }$ & -0.434$^{\rm }$ & 0.417$^{\rm }$&0.800$^{\rm }$&\protect\cite{b27,b28}\\
\hline
~8 & 0.261$\pm$0.004 & 1.643$\pm$0.127 &  0.061$\pm$0.019 & 0.851$\pm$0.083 & -0.553$\pm$0.125  & 0.416$\pm$0.030 & 0.821$\pm$0.014 & \\
 & & 1.638(1.612) & 0.0247(0.0246) & 0.829(0.825) &  & & 0.810(0.808)&\protect\cite{b2} \\
 & 1.305$^{\rm }$ & 1.637$^{\rm }$ & 0.027$^{\rm }$& 0.830$^{\rm }$ & -0.489$^{\rm }$ & 0.426$^{\rm }$&0.808$^{\rm }$&\protect\cite{b27,b28}\\
\hline
~9 & 0.243$\pm$0.004  & 1.680$\pm$0.140  & 0.059$\pm$0.019 & 0.869$\pm$0.089 & -0.608$\pm$0.134   & 0.422$\pm$0.032   & 0.825$\pm$0.012  &  \\
   &  & 1.680(1.643) & 0.0233(0.0233) & 0.850(0.841)&  & &0.817(0.815)&\protect\cite{b2} \\
   & 1.289$^{\rm }$ & 1.669$^{\rm }$& 0.025$^{\rm }$ & 0.845$^{\rm }$ & -0.536$^{\rm }$ & 0.433$^{\rm }$&0.815$^{\rm }$&\protect\cite{b27,b28}\\
\hline
10 & 0.226$\pm$0.003  & 1.713$\pm$0.150 & 0.057$\pm$0.017 & 0.884$\pm$0.093  & -0.654$\pm$0.140  & 0.426$\pm$0.032 &0.828$\pm$0.010 & \\
 &  &1.713(1.670)& 0.0216(0.0220) & 0.866(0.854) &  & &0.824(0.822)&\protect\cite{b2} \\
 & 1.275$^{\rm }$& 1.697$^{\rm }$& 0.024$^{\rm }$ & 0.859$^{\rm }$ & -0.576$^{\rm }$ & 0.440$^{\rm }$&0.822$^{\rm }$&\protect\cite{b27,b28}\\
\hline
12 & 0.199$\pm$0.003   & 1.765$\pm$0.163  &  0.054$\pm$0.015  & 0.908$\pm$0.098 & .0726$\pm$0.147  & 0.480$\pm$0.032  & 0.836$\pm$ 0.007 & \\
 &  &1.763(1.716) & 0.0190(0.0198) & 0.890(0.877) &  & &0.838(0.835)&\protect\cite{b2} \\
 & 1.249$^{\rm }$& 1.743$^{\rm }$& 0.021$^{\rm }$ & 0.881$^{\rm }$ & -0.643$^{\rm }$ & 0.450$^{\rm }$ &0.836$^{\rm }$&\protect\cite{b27,b28}\\
\hline
14 & 0.178$\pm$0.002  & 1.804$\pm$0.170 & 0.048$\pm$0.012 & 0.925$\pm$0.099  & -0.777$\pm$0.148   & 0.486$\pm$0.031 & 0.843$\pm$0.006  &    \\
 & &1.795(1.750) & 0.0169(0.0178) & 0.905(0.894) &  & &0.851(0.849)&\protect\cite{b2}    \\
 & 1.227$^{\rm }$& 1.779$^{\rm }$& 0.019$^{\rm }$ & 0.898$^{\rm }$ & -0.693$^{\rm }$& 0.457$^{\rm }$&0.849$^{\rm }$&\protect\cite{b27,b28}\\
\hline
16 & 0.160$\pm$0.002   & 1.833$\pm$ 0.172 & 0.042$\pm$0.010   & 0.938$\pm$0.097 & -0.814$\pm$0.146  & 0.490$\pm$0.030  & 0.850$\pm$0.004  &\\
 &  & 1.822(1.779) & 0.0152(0.0161) & 0.918(0.907) &  & &0.862(0.860) &\protect\cite{b2}\\
 & 1.208$^{\rm }$ & 1.807$^{\rm }$& 0.017$^{\rm }$& 0.911$^{\rm }$ & -0.732$^{\rm }$& 0.463$^{\rm }$&0.861$^{\rm }$&\protect\cite{b27,b28}\\
\hline
18 & 0.146$\pm$0.001   &  1.856$\pm$0.171 & 0.038$\pm$0.008  & 0.946$\pm$0.095 & -0.840$\pm$0.142  & 0.492$\pm$-0.028  & 0.856$\pm$0.003  &\\
 & &1.845(1.803) & 0.0148(0.0137) & 0.929(0.918) && &0.873(0.869) &\protect\cite{b2}\\
 & 1.191$^{\rm }$ & 1.829$^{\rm }$ & 0.015$^{\rm }$ & 0.921$^{\rm }$ & -0.764$^{\rm }$ & 0.468$^{\rm }$&0.871$^{\rm }$&\protect\cite{b27,b28}\\
\hline
20 & 0.134$\pm$0.001  & 1.873$\pm$0.168  & 0.033$\pm$0.006  & 0.953$\pm$0.091  & -0.861$\pm$0.137 & 0.493$\pm$0.026   & 0.862$\pm$0.002   &\\
 &        &1.864(1.822) & 0.0125(0.0135) & 0.938(0.927)& &&0.883(0.878)&\protect\cite{b2}\\
 & 1.177$^{\rm }$ & 1.847$^{\rm }$ & 0.014$^{\rm }$ & 0.930$^{\rm }$ & -0.789$^{\rm }$ & 0.471$^{\rm }$&0.880$^{\rm }$&\protect\cite{b27,b28}\\
\hline
24 &0.114$\pm$0.001  &  1.898$\pm$0.158 & 0.027$\pm$0.004 & 0.963$\pm$ 0.084 & -0.889$\pm$0.126  & 0.429$\pm$0.023  & 0.873$\pm$0.001  & \\
 & &1.890(1.850) & 0.0106(0.0116) & 0.950(0.939) & &&0.900(0.894)&\protect\cite{b2} \\
 & 1.154$^{\rm }$& 1.874$^{\rm }$ & 0.012$^{\rm }$ & 0.942$^{\rm }$ & -0.827$^{\rm }$ & 0.477$^{\rm }$&0.896$^{\rm }$  &\protect\cite{b27,b28}\\
\hline
28 & 0.100$\pm$0.001  & 1.915$\pm$0.148 & 0.023$\pm$0.002 & 0.969$\pm$ 0.077 & -0.907$\pm$0.116  &0.427$\pm$0.021  & 0.882$\pm$ 0.000 & \\
&& 1.909(1.871)& 0.009232(0.01010)& 0.959(0.949) & &&0.913(0.906)&\protect\cite{b2} \\
 & 1.136$^{\rm }$ & 1.893$^{\rm }$& 0.010$^{\rm }$& 0.951$^{\rm }$ & -0.854$^{\rm }$ & 0.481$^{\rm }$&0.909$^{\rm }$&\protect\cite{b27,b28}\\
\comment{\hline
32 &0.088$\pm$ 0.000 & 1.926$\pm$0.138 & 0.020$\pm$ 0.001& 0.973$\pm$0.071 & -0.919$\pm$0.106  & 0.425$\pm$0.018  & 0.890$\pm$ 0.000 & \\
&& 1.920(1.887) & 0.00814(0.00895)& 0.964(0.955)& &&0.924(0.915)&\protect\cite{b2} \\
 & 1.122$^{\rm }$ & 1.908$^{\rm }$ & 0.009$^{\rm }$& 0.958$^{\rm }$& -0.875$^{\rm }$& 0.483$^{\rm }$&0.919$^{\rm }$&\protect\cite{b27,b28}\\}
\hline
$\infty$ &0 & 2/($D$-2) & 0&1/($D$-2) &($D$-4)/($D$-2) & 1/2  & 4-$D$ \\
\hline
\end{tabular} }

\begin{figure}[tbhp]
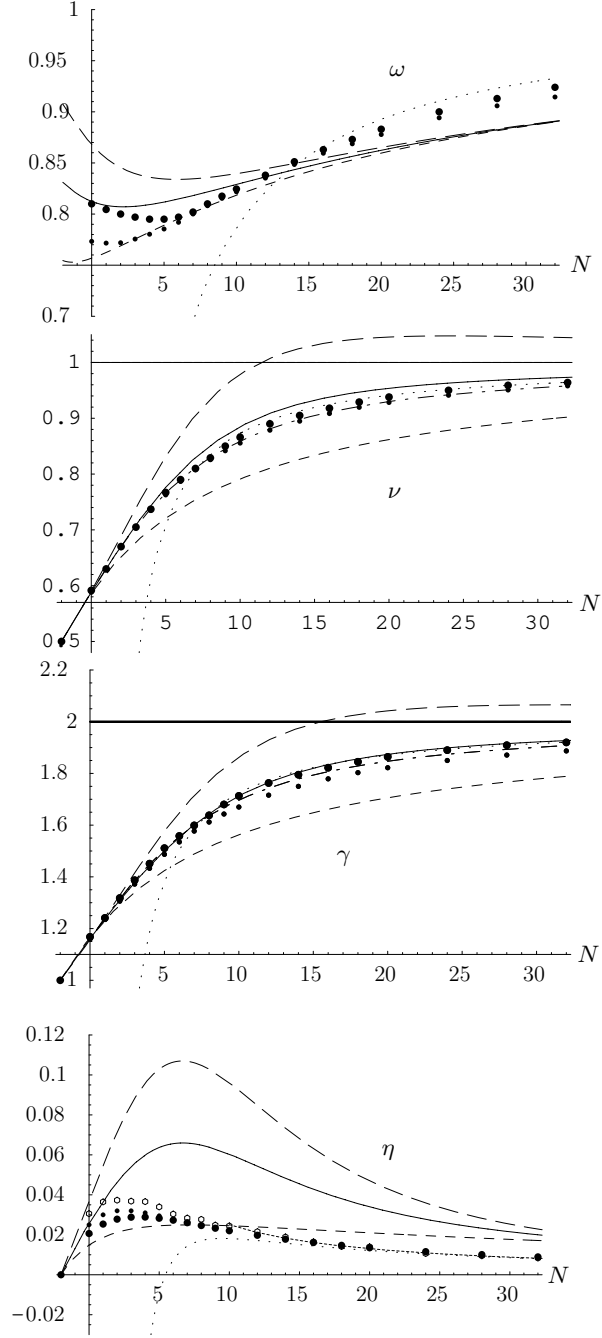

\vspace{2.7cm}
\phantom{,xxx}~$\!\!\!$\input omln.tps
\vspace{2.5cm}
~\input nuln.tps
\vspace{2.7cm}
\input gamln.tps
\vspace{2.7cm}
\input etaln.tps
\caption[]
{Solid curves show our fourth-order approximations to $\omega,\;\nu,\;
\gamma,\;\eta$. Short-dashed is second-, long-dashed curve is third-order
approximation. Thin dots show sixth-order approximation of the textbook
\cite{b2}, fat dots the extrapolations to infinite order. The dash-dotted
lines in the second and third figures are interpolations to the Pad\'e-Borel
resummations of \cite{b27,b28} (where $ \omega$ was not calculated).
Their data for $ \eta$ scatter too much to be represented in this way--they
are indicated by small circles in the fourth figure. The dotted curves show
$1/N$~-expansions of all four quantities.  Note that our results lie closer
to these than those of \aut{S.A. Antonenko} and \aut{A.I. Sokolov}. The solid
$\eta$-curve was calculated in the textbook \cite{b2} (see Fig.~20.2). The
exact large-$N$ limits are $\omega _{N=\infty}=4-D$, $\nu _{N=\infty}=1/(D-2)$,
$ \gamma _{N=\infty}=2/(D-2)$, and $\eta _{N=\infty}=0$. The exact values at
$N=-2$ are $\nu _{N=-2}=1/2$, $ \gamma _{N=-2}=1$, and $\eta _{N=-2}=0$ for
all $D$.
}
\label{@}\end{figure}


\vskip 5mm

\begin{thebibliography}{99}

\bibitem{b2}
H. Kleinert and V. Schulte-Frohlinde, {\it Critical Properties of
$\vp^4$ Theories} (World Scientific, Singapore, 2001).

\bibitem{b1}
J. Zinn-Justin, {\it Quantum Field Theory and Critical Phenomena}
(Oxford University, Oxford, 1996).

\bibitem{b3}
H. Kleinert, Phys. Rev. D {\bf 57}, 2264 (1998).

\bibitem{b4}
H. Kleinert, Phys. Lett. B {\bf 434}, 74 (1998).

\bibitem{b5}
H. Kleinert, Phys. Rev. D {\bf 60}, 085001 (1999), hep-th/9812197.

\bibitem{b6}
H. Kleinert and B. Van den Bossche, Phys. Rev. E {\bf 63}, 056113 (2001).

\bibitem{b7}
B. Simon, Bull. Am. Math. Soc. {\bf 24}, 303 (1991).

\bibitem{b8}
H. Kleinert, {\it Path Integrals in Quantum Mechanics, Statistics,
Polymer Physics, and Financial Markets} (World Scientific, Singapore,
2003).

\bibitem{b9}
V.I. Yukalov, Physica A {\bf 167}, 833 (1990).

\bibitem{b10}
V.I. Yukalov, Phys. Rev. A {\bf 42}, 3324 (1990).

\bibitem{b11}
V.I. Yukalov, J. Math. Phys. {\bf 32}, 1235 (1991).

\bibitem{b12}
V.I. Yukalov, J. Math. Phys. {\bf 33}, 3994 (1992).

\bibitem{b13}
V.I. Yukalov and E.P. Yukalova, Physica A {\bf 225}, 336 (1996).

\bibitem{b14}
V.I. Yukalov and E.P. Yukalova, Ann. Phys. (N.Y.) {\bf 277}, 219 (1999).

\bibitem{b15}
V.I. Yukalov and E.P. Yukalova, Chaos Solitons Fractals {\bf 14}, 839
(2002).

\bibitem{b16}
B. Hamprecht and H. Kleinert, Phys.Rev. D {\bf 68}, 065001 (2003),
(hep-th/0302116)


\bibitem{b17}
H.H.H. Homeier, Acta Appl. Math. {\bf 61}, 133 (2000).

\bibitem{b18}
W. Janke and H. Kleinert, Phys. Rev. Lett. {\bf 75}, 2787 (1995).

\bibitem{b19}
H. Kleinert, Phys. Lett. B {\bf 463}, 69 (1999), (cond-mat/9906359)

\bibitem{b20}
V.I. Yukalov, Moscow Univ. Phys. Bull. {\bf 31}, 10 (1976).

\bibitem{b21}
R. Hasson and D. Richards, J. Phys. B {\bf 34}, 1805 (2001).

\bibitem{b22}
V.I. Yukalov, Phys. Lett. A {\bf 284}, 91 (2001).

\bibitem{b23}
V.I. Yukalov, Physica A {\bf 291}, 255 (2001).

\bibitem{b24}
F.J. Wegner, Phys. Rev. B {\bf 5}, 4529 (1972).

\bibitem{b25}
J.B. Kogut, Rev. Mod. Phys. {\bf 51}, 659 (1979).

\bibitem{b26}
V.I. Yukalov and A.S. Shumovsky, {\it Lectures on Phase Transitions}
(World Scientific, Singapore, 1990).

\bibitem{b27}
S.A. Antonenko and A.I. Sokolov, Phys. Rev. E {\bf 51}, 1894 (1995).

\bibitem{b28}
S.A. Antonenko and A.I. Sokolov, Phys. Solid State {\bf 40}, 1169
(1998).

\bibitem{b29}
G.A. Baker Jr., B.G. Nickel, and D.I. Meiron, Phys. Rev. B {\bf 17},
1365 (1978).

\bibitem{b30}
J.G. Le Guillou and J. Zinn-Justin, Phys. Rev. Lett. {\bf 39}, 95
(1977).

\bibitem{b31}
J.G. Le Guillou and J. Zinn-Justin, Phys. Rev. B {\bf 21}, 3976
(1980).

\bibitem{b32}
R. Guida and J. Zinn-Justin, J. Phys. A {\bf 31}, 8103 (1998).

\bibitem{b33}
K.G. Chetyrkin, S.G. Gorishny, S.A. Larin, and F.V. Tkachov, Phys. Lett.
B {\bf 132}, 351 (1983).

\bibitem{b34}
S.G. Gorishny, S.A. Larin, and F.V. Tkachov, Phys. Lett. A {\bf 101},
120 (1984).

\bibitem{b35}
H. Kleinert, J. Neu, V. Shulte-Frohlinde, K.G. Chetyrkin, and S.A. Larin,
Phys. Lett. B {\bf 272}, 39 (1991).

\bibitem{b36}
J.G. Le Guillou and J. Zinn-Justin, J. Phys. Lett. {\bf 46}, 137 (1985).

\bibitem{b37}
J. Honkonen and M. Nalimov, Phys. Lett. B {\bf 459}, 582 (1999).

\bibitem{b38}
D.F. Litim, Int. J. Mod. Phys. A {\bf 16}, 2081 (2001).

\bibitem{b39}
D.F. Litim, J. High Energy Phys. {\bf 11}, 059 (2001).

\bibitem{b40}
J. Honkonen, M. Komarova, and M. Nalimov, Acta Phys. Slov. {\bf 52},
303 (2002).

\bibitem{b41}
V.I. Yukalov and S. Gluzman, Phys. Rev. E {\bf 58}, 1359 (1998).

\bibitem{b42}
D.P. Landau, J. Magn. Magn. Mater. {\bf 200}, 231 (1999).

\bibitem{b43}
J. Zinn-Justin, Phys. Rep. {\bf 344}, 159 (2001).

\bibitem{b44}
A. Pelissetto and E. Vicari, Phys. Rep. {\bf 368}, 549 (2002).

\bibitem{b45}
J.A. Lipa, D.R. Swanson, J. Nissen, Z.K. Geng, P.R. Williamson,
D.A. Stricker, T.C.P. Chui, U.E. Israelson, and M. Larson, Phys. Rev.
Lett. {\bf 84}, 4894 (2000).

\bibitem{b46}
J.A. Lipa, J.A. Nissen,  D.A. Stricker, D.R. Swanson, and T.C.P. Chui,
Phys. Rev. B {\bf 68}, 174518 (2003).

\end{thebibliography}
\end{document}